\documentclass[cernpreprint, atlasdraft=false, texlive=2023, UKenglish, texmf, orcidlogo]{atlasdoc}
 
\usepackage{atlaspackage}
\usepackage{atlasbiblatex}
 
\usepackage{atlasphysics}

\graphicspath{{logos/}{figures/}}
 
\usepackage{ANA-TOPQ-2018-44-PAPER-defs}
\AtlasTitle{Observation of Single-Top-Quark Production in Association with a Photon Using the ATLAS Detector}
 
\AtlasAbstract{
This Letter reports the observation of single top quarks produced together with a photon, which directly probes the electroweak coupling of the top quark. The analysis uses 139~fb$^{-1}$ of 13~TeV proton--proton collision data collected with the ATLAS detector at the Large Hadron Collider. Requiring a photon with transverse momentum larger than 20~GeV and within the detector acceptance, the fiducial cross section is measured to be $688\,\pm\,23 \, \mathrm{(stat.)} \, ^{+75}_{-71} \, \mathrm{(syst.)}$~fb, to be compared with the standard model prediction of $515\,^{+36}_{-42}$~fb at next-to-leading order in QCD.
}
 
\author{The ATLAS Collaboration}
 
\PreprintIdNumber{CERN-EP-2022-229}

\AtlasJournalRef{\PRL 131 (2023) 181901}
\AtlasDOI{10.1103/PhysRevLett.131.181901}

\hypersetup{pdftitle={ATLAS document},pdfauthor={The ATLAS Collaboration}}
 
\begin{document}
 
\maketitle

Measurements of rare associated-production processes of the top quark ($t$) are crucial in probing the top quark's electroweak couplings, which are fundamental quantities of the standard model (SM). While pair production ($t\bar{t}$) has been observed in association with a Higgs boson~\cite{HIGG-2018-13,CMS-HIG-17-035}, $W$ boson~\cite{CMS-TOP-17-005}, $Z$ boson~\cite{TOPQ-2016-11,CMS-TOP-17-005}, or photon ($\gamma$)~\cite{TOPQ-2012-07}, single-top-quark production has so far only been observed in association with a $Z$~\cite{TOPQ-2018-01,CMS-TOP-18-008} or $W$ boson~\cite{TOPQ-2012-20,CMS-TOP-12-040}. These processes play a crucial role in constraining nonresonant contributions from physics beyond the SM (BSM), parametrized in the framework of the SM effective field theory~\cite{Buckley:2015lku,Hartland:2019bjb,Durieux:2019rbz,Brivio:2019ius,Miralles:2021dyw}. This Letter reports the observation of single-top-quark production in association with a photon in the dominant $t$-channel mode with the ATLAS detector~\cite{PERF-2007-01} at the Large Hadron Collider (LHC). The full 13~\TeV\ proton--proton ($pp$) dataset is used, corresponding to an integrated luminosity of 139~\ifb~\cite{ATLAS-CONF-2019-021}. The CMS Collaboration previously reported evidence for this process using 35.9~fb$^{-1}$ of $pp$ data collected at 13~\TeV~\cite{CMS-TOP-17-016}.
 
In single-top-quark production, a photon can be radiated from any of the charged particles in the initial and final states, but the radiation before the top-quark decay is of particular interest. This process, in the following denoted as $\sigprod$, where $q$ stands for the additional quark produced in the $t$ channel, represents a direct probe of the top--photon coupling and offers sensitivity to BSM contributions comparable to established probes, e.g., photon-associated top-quark pair production ($\ttbar\gamma$)~\cite{Fael:2013ira,Etesami:2016rwu}. This Letter only considers semileptonic top-quark decays in $\sigprod$ production as they provide better sensitivity than hadronic decays. An example Feynman diagram is shown in Figure~\ref{fig:Feynman}. The signature of this process consists of a photon, an electron or muon ($\ell$), missing transverse momentum (\met) from the neutrino ($\nu$), a $b$ jet from the top-quark decay, and a forward jet characteristic of $t$-channel production. The jet arising from the second $b$ quark from gluon splitting is often not $b$ tagged because of its low transverse momentum and forward direction. The photon can also be radiated from the top quark's charged decay products, called the $\sigdecay$ process.
 
\begin{figure}[h]
\centering
\includegraphics[width=0.5\textwidth]{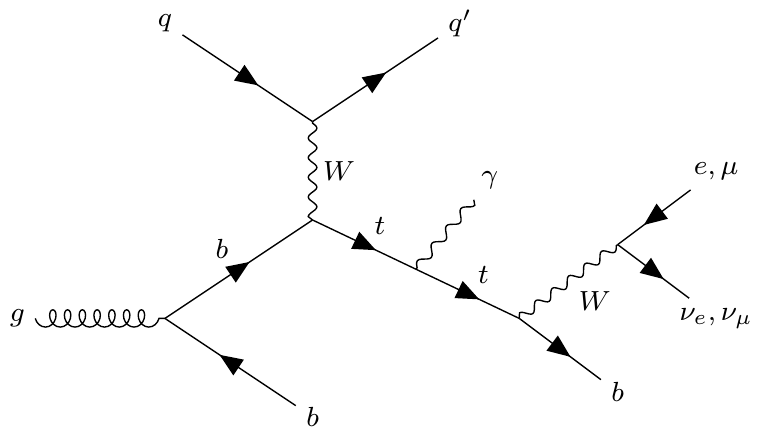}
\caption{Representative Feynman diagram at leading order in \alphas for $\sigprod$ production with semileptonic top-quark decay.}
\label{fig:Feynman}
\end{figure}
 
Two cross section measurements are performed in fiducial phase spaces. A measurement of the cross section of the combined $\sigprod$ and $\sigdecay$ processes is performed in a fiducial phase space at stable-particle level. In addition, the cross section for $\sigprod$ production alone is measured in a fiducial phase space that allows for a direct comparison with fixed-order predictions where, in contrast to the phase space at stable-particle level, parton showering and hadronization are not considered.
 
The ATLAS detector is a multipurpose particle physics detector with cylindrical geometry\footnote{ATLAS uses a right-handed coordinate system with its origin at the nominal interaction point (IP) in the center of the detector and the \(z\) axis along the beam pipe. The \(x\) axis points from the IP to the center of the LHC ring, and the \(y\) axis points upwards. Cylindrical coordinates \((r,\phi)\) are used in the transverse plane, \(\phi\) being the azimuthal angle around the \(z\) axis. The pseudorapidity is defined in terms of the polar angle \(\theta\) as \(\eta = -\ln \tan(\theta/2)\). Angular distance is measured in units of \(\Delta R \equiv \sqrt{(\Delta\eta)^{2} + (\Delta\phi)^{2}}\).}. It consists of an inner tracker surrounded by a superconducting solenoid, sampling electromagnetic and hadronic calorimeters, and a muon spectrometer with three toroidal superconducting magnets with eight coils each. A two-level trigger system is used to select events for storage. Events used in this analysis were selected online by sets of single-electron or single-muon triggers with their lowest transverse-momentum (\pt) thresholds being 20--26~\GeV, depending on the data-taking year~\cite{TRIG-2016-01,TRIG-2018-05,TRIG-2018-01}. An extensive software suite~\cite{ATL-SOFT-PUB-2021-001} is used in data simulation, in the reconstruction and analysis of real and simulated data, in detector operations, and in the trigger and data acquisition systems of the experiment.
 
The selection of the primary proton-interaction vertex as well as the reconstruction and identification of electrons, muons, photons, and jets follow Ref.~\cite{TOPQ-2019-19}, with the difference that jets are considered with $|\eta|$ up to 4.5, which accounts for the forward jet from the $t$-channel production. The definition of \met\ and the $b$-tagging algorithm (DL1r~\cite{FTAG-2019-07-tqg}) for the jets are identical to those in Ref.~\cite{TOPQ-2019-19}.
 
Signal and most background processes are estimated with Monte Carlo (MC) simulations, which include pileup effects. The full \GEANT-based~\cite{Agostinelli:2002hh} ATLAS detector simulation is used~\cite{SOFT-2010-01} with corrections applied for the different reconstructed objects~\cite{EGAM-2018-01,MUON-2018-03,FTAG-2018-01,JETM-2018-05,PERF-2014-03,PERF-2016-07}.
 
Two signal regions (SRs) are defined, based on the presence or absence of the forward jet (fj) in the event. In both SRs, the presence of one photon, one electron or muon matched to a trigger object, one tight $b$-tagged jet, no additional loose $b$-tagged jets, and $\met>30~\GeV$ is required. In addition, the 0fj~SR ($\geq$1fj~SR) must contain no (at least one) forward jet with $2.5 < |\eta| < 4.5$. The tight and loose operating points of the $b$-tagging algorithm correspond to efficiencies of 70\% and 85\% and to misidentification rates of 8\% (0.2\%) and 35\% (2.5\%) for $c$ jets (light-flavor jets), estimated in $t\bar{t}$ MC simulations. In both SRs, the electron--photon invariant mass must be outside the range 80--100~\GeV\ to suppress the $Z\rightarrow ee$ contribution with an electron misidentified as a photon.
 
The measured combined $\sigprod$ and $\sigdecay$ rate is unfolded to a fiducial phase space that is defined at stable-particle level, where ``stable'' refers to lifetimes larger than 30~ps, and is translated into a fiducial cross section times branching ratio. The definitions of photons, photon isolation, electrons, muons, jets, and $b$-tagged jets at stable-particle level follow Ref.~\cite{TOPQ-2017-14}, except that jets are considered with $|\eta|$ up to 4.9. This phase space is defined close to the SRs by requiring one electron or muon with $\pt > 25~\GeV$ and $|\eta| < 2.5$, at least one photon with $\pt > 20~\GeV$ and $|\eta| < 2.37$, at least one $b$-tagged jet with $\pt > 25~\GeV$ and $|\eta| < 2.5$, and at least one neutrino that is not produced in a hadron decay. Jets are required to be separated by more than $\Delta R = 0.4$ from any lepton and isolated photon. No photon must be within $\Delta R = 0.4$ of any jet or lepton. The SM fiducial cross section at stable-particle level times branching ratio, where the branching ratio of the semileptonic top-quark decay is denoted by $\mathcal{B}\left(t\rightarrow \ell\nu b\right)$, is calculated at next-to-leading order (NLO) in QCD using the signal samples for $\sigprod$ and $\sigdecay$ defined below: $\sigma_{\sigprod} \times \mathcal{B}\left(t\rightarrow \ell\nu b\right) + \sigma_{\sigdecay} = 217\,^{+27}_{-15}$~fb. The branching ratio for $t\rightarrow \ell\nu b$ is set to 32.46\%, consistent with the value in the signal MC samples. The uncertainty includes variations of the parton distribution functions (PDFs) and of the scales, uncertainties in the parton shower model, the choice of matrix-element generator, the modeling of initial- and final-state radiation, and a 20\% uncertainty in the $\sigdecay$ process normalization (cf.\ the Appendix). The $\sigdecay$ process constitutes $\approx 20\%$ of the events in the fiducial region.
 
Additionally, the measured $\sigprod$ rate is unfolded to a fiducial phase space and is translated into a fiducial cross section times branching ratio. The phase space is defined before hadronization and parton showering by requiring at least one photon with $\pt > 20~\GeV$ and $|\eta|<2.37$ that must be Frixione isolated~\cite{Frixione:1998jh} with a chosen isolation radius of $\Delta R = 0.2$. Following Ref.~\cite{Pagani:2021iwa}, the fixed-order SM fiducial cross section times branching ratio is calculated with \MGNLO~\cite{Alwall:2014hca} at NLO in QCD as $\sigma_{\sigprod}\times \mathcal{B}\left(t\rightarrow \ell\nu b\right) = 515\,^{+36}_{-42}$~fb. The cross section calculation uses the five-flavor scheme, with $b$ quarks included in the proton. Renormalization and factorization scales as well as the PDF set are chosen as in Ref.~\cite{Pagani:2021iwa}. The uncertainties are estimated from scale and PDF variations and from a comparison with the corresponding calculation in the four-flavor scheme (no third-generation quarks in the proton)~\cite{Pagani:2020mov,Pagani:2021iwa}.
 
The $\sigprod$ process was simulated in the four-flavor scheme at NLO in QCD with \MGNLO\ using the NNPDF3.0~\cite{Ball:2014uwa} PDF set and \MADSPIN~\cite{Artoisenet:2012st} for $t\rightarrow Wb\rightarrow\ell\nu b$ decay. Photons must be Frixione isolated and have $\pt > 10~\GeV$ and $|\eta| < 5.0$. Renormalization and factorization scales were set to $\frac{1}{2}\sum_i \sqrt{m_i^2 + p_{\mathrm{T}, i}^2}$, where the sum is over all final-state particles before the top-quark decay. \Pythia[8]~\cite{Sjostrand:2014zea} was used for parton showering and hadronization. \Pythia[8] always used the leading-order (LO) NNPDF2.3 PDF set~\cite{Ball:2012cx}, the A14 tune~\cite{ATL-PHYS-PUB-2014-021}, and \EVTGEN~\cite{Lange:2001uf}. The $\sigdecay$ process was simulated via single-top-quark production in the $t$ channel (without photon radiation) using \POWHEG~\cite{Frederix:2012dh} in the four-flavor scheme at NLO with the NNPDF3.0 PDF set, interfaced to \Pythia[8] and \MADSPIN\ for the semileptonic top-quark decay. Photon radiation in the decay was treated by the parton-shower simulation. Initially, the $\sigprod$ process is normalized to the cross section at NLO in QCD obtained with \MGNLO\ and the same settings used for the sample production, and the $\sigdecay$ process is normalized to the production cross section at NLO in QCD~\cite{Kant:2014oha,Aliev:2010zk}. The overlap between the $\sigprod$ and $\sigdecay$ samples is removed using kinematic information about the generated particles (cf.\ the Appendix).
 
The most important background processes with prompt photons are $t\bar{t}\gamma$ production, which refers to photon radiation in $t\bar{t}$ production and $t\bar{t}$ production with radiative decay ($\decay$ and $t\rightarrow q\bar{q}^\prime b\gamma$), and $W\gamma$+jets production. Their contribution to the SRs is estimated using MC simulations (cf.\ the Appendix), normalized to data in dedicated control regions (CRs) enriched in $t\bar{t}\gamma$ and $W\gamma$+jets events. The CRs are inclusive in forward jets and the same selection criteria as for the SRs are used, except for the $b$-tagging requirements. In the $t\bar{t}\gamma$~CR, an additional loose $b$-tagged jet must be present to account for the second $b$ quark in this process. In the $W\gamma$~CR, there must be at least one loose $b$-tagged jet and no tight $b$-tagged jets. This suppresses contamination of the CR with processes with one or several $b$ quarks in the final state. Other background contributions that are modeled by MC simulations (cf.\ the Appendix) are the following production processes: $Z\gamma$+jets, $t\bar{t}$, single top quark, $W$/$Z$+jets, and diboson. The events from these MC samples, apart from $Z\gamma$+jets events, are categorized into events with prompt photons (``other prompt $\gamma$''), electrons misidentified as photons ($\efake$), and hadrons misidentified as photons ($\hfake$). All background MC samples use the same setup as in Ref.~\cite{TOPQ-2019-19}. An additional small background contribution arises from events with fake leptons, i.e., other objects that are misidentified as electron or muon, and is estimated from data using the asymptotic matrix method with loosened lepton criteria~\cite{D0:2007oab,EGAM-2019-01}.
 
The MC predictions for background processes with $\efake$ fakes, most notably dileptonic $t\bar{t}$ events, are corrected by comparing the $\efake$ probability in data and MC simulation using $Z\rightarrow e^+e^-$ events (see Supplemental Material~\cite{supp}). Events with $\met < 30~\GeV$ and no $b$-tagged jet are selected if the invariant mass of either an $e^+e^-$ pair or an $e\gamma$ pair is close to the $Z$-boson mass, where the photon in the latter case is likely from $\efake$. Data-to-MC corrections are derived as functions of the photon $\eta$ and the different types of photon reconstruction~\cite{EGAM-2018-01}. No strong dependence of the corrections on the photon $\pt$ is found. The corrections are validated by comparing data with the prediction in a region with $\met<30~\GeV$ and at least one $b$-tagged jet.
 
The MC predictions for background processes with $\hfake$ fakes, mostly lepton+jets $t\bar{t}$ events, are also corrected using data~\cite{supp}. Selections with partially inverted photon-identification and/or inverted photon-isolation criteria are used, respectively, to define regions that are kinematically close to the analysis regions but enriched in events with $\hfake$. Considering the low correlation between the identification and isolation criteria, the ABCD method (see, for example, Ref.~\cite{CDF:1990kbl}) is used to estimate the number of $\hfake$ events in the analysis regions. This residual small correlation is taken from MC simulations and is corrected for in the estimate. The $\hfake$ rate estimate is performed in two bins of photon $\pt$ and as a function of photon reconstruction types and $\eta$, and is used to correct the overall normalization of the contribution from $\hfake$ events.
 
According to these signal and background predictions, the signal fraction [``$\sigprod$'' plus ``$\sigdecay$''] in the 0fj ($\geq$1fj) SR is 5\% (10\%) and the main backgrounds are $t\bar{t}\gamma$ with 29\% (34\%), $\efake$ fake-photon events with 24\% (25\%), $W\gamma$+jets with 20\% (12\%), and $\hfake$ fake events with 7\% (7\%). Smaller backgrounds originate from $Z\gamma$+jets, the other prompt $\gamma$ contribution, and events with fake leptons.
 
Uncertainties in the photon identification~\cite{EGAM-2018-01,PERF-2017-02} and isolation efficiencies~\cite{EGAM-2018-01} are considered, as are those in the electron and muon trigger, reconstruction, identification, and isolation efficiencies~\cite{EGAM-2018-01,PERF-2017-01,PERF-2015-10}; the photon and electron energy~\cite{EGAM-2018-01} and muon momentum scale and resolution~\cite{PERF-2015-10}; the jet pileup rejection~\cite{PERF-2014-03}, and jet energy scale~\cite{JETM-2018-05,PERF-2016-04}, resolution~\cite{JETM-2018-05}, and $b$-tagging efficiency~\cite{FTAG-2018-01,FTAG-2019-02,FTAG-2020-08}; and the \met\ reconstruction~\cite{PERF-2016-07}.
 
Uncertainties in the inclusive cross sections and in the modeling (scale variations, comparisons of generator setups, etc.) of the different processes are considered (cf.\ the Appendix). Since the analysis includes CRs for the $t\bar{t}\gamma$ and $W\gamma$+jets processes, their normalization is estimated directly from data.
 
Uncertainties in the $\efake$ corrections are estimated by varying the background contributions, the $Z$-boson MC modeling, the $Z$-boson mass range, and the photon energy scale (see Supplemental Material~\cite{supp}). Uncertainties in the $\hfake$ corrections originate from the statistical uncertainties, the limited number of MC events, contributions from non-$\hfake$ events, and variations of the correlation between the inverted identification and isolation criteria~\cite{supp}.
 
Neural networks (NNs) are trained to separate the signal from the background in the SRs. Keras~\cite{chollet2015keras} with the TensorFlow~\cite{tensorflow2015-whitepaper} back end is used with binary cross-entropy as the loss function. The NN output nodes use a sigmoid activation function. In the 0fj and $\geq$1fj~SRs, 12 and 15 input variables are used, respectively. These comprise individual kinematic properties (\pt and/or $\eta$) of the photon, the lepton, the $b$-tagged jet, and the highest-$\pt$ forward jet, kinematic combinations (scalar \pt\ sum, invariant and transverse masses, angular distances, transverse momentum, energy) of these objects, and the $\met$, as well as the lepton type and the $b$-tagging properties of the $b$-tagged jet~\cite{FTAG-2018-01}. The top quark is reconstructed from the $b$-tagged jet, the lepton, and the $\met$. The top-quark mass is the NN input variable giving the largest separation in both SRs as it separates $\sigprod$ from backgrounds without a top quark, top-quark events with $\decay$, as well as top-quark pair production where the chosen objects are less likely associated with the same top-quark decay. Figure~\ref{fig:input} shows this variable in the $W\gamma$~CR as an example, illustrating that the data are described by the MC simulation within the uncertainties.
 
\begin{figure}[h]
\centering
\includegraphics[width=0.4\textwidth]{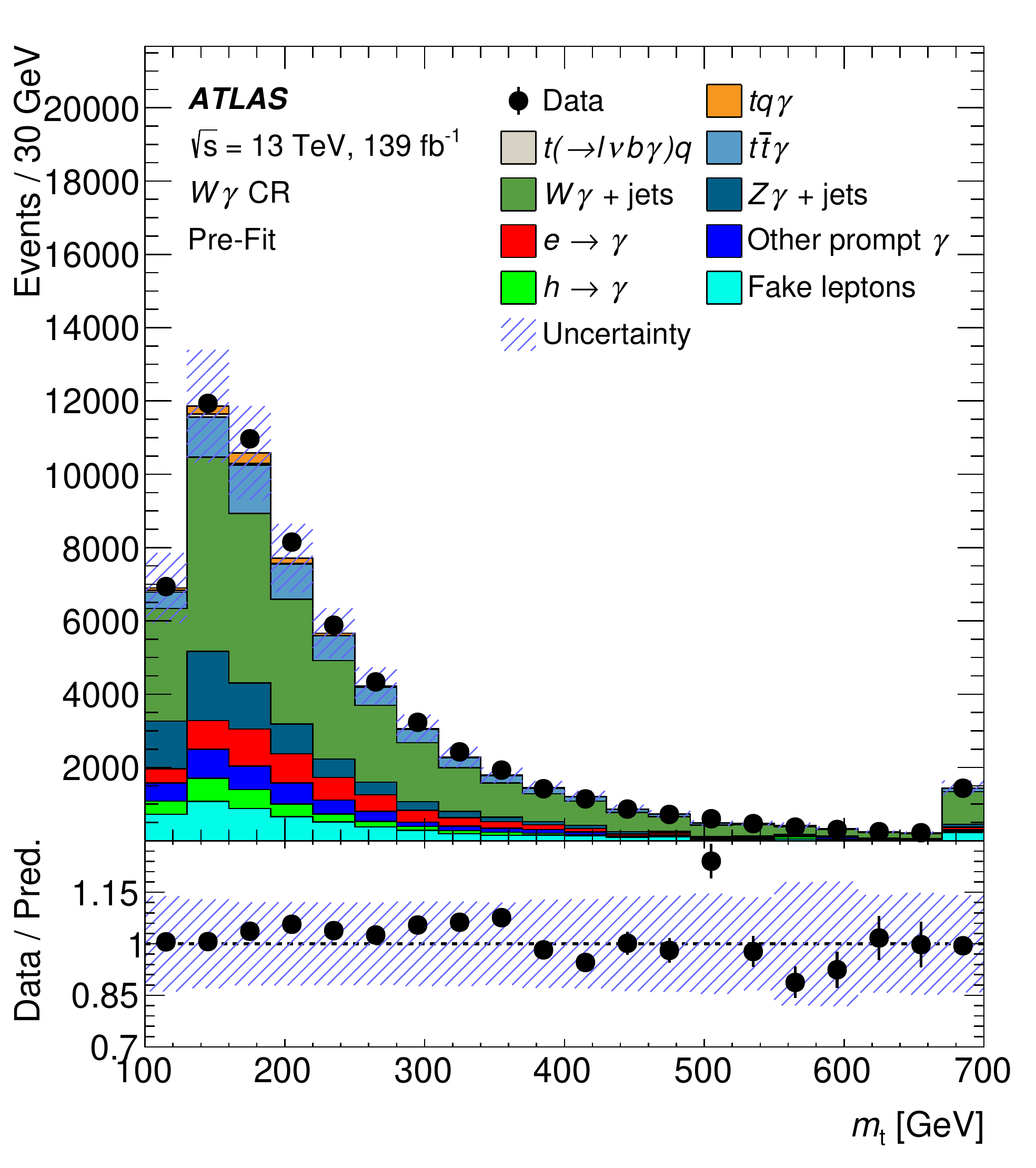}
\caption{Distribution of the reconstructed top-quark mass in the $W\gamma$~CR before the profile-likelihood fit. The hashed band represents the uncertainties and the first and last bins include the underflow and overflow.}
\label{fig:input}
\end{figure}
 
To test for the presence of $\sigprod$ production and measure the signal cross sections, a profile-likelihood fit using asymptotic formulas~\cite{Cowan:2010js} is performed simultaneously in the SRs and CRs with systematic uncertainties treated as nuisance parameters. The uncertainty due to the limited number of MC events is included~\cite{Barlow:1993dm}. In the 0fj ($\geq$1fj) SRs, the 0fj ($\geq$1fj) NN output distributions are used in the fit. In the $t\bar{t}\gamma$~CR, the 0fj ($\geq$1fj) NN output is used for events with no (at least one) forward jet, and the inclusive event yield is used in the $W\gamma$~CR. The $t\bar{t}\gamma$ and $W\gamma$+jets normalizations are free parameters of the fit. The result of the fit is shown in Figure~\ref{fig:postfit}. The predicted sum of all backgrounds is not compatible with the data. The observed (expected) significance of the $\sigprod$ signal is 9.3$\sigma$ (6.8$\sigma$). The fitted $t\bar{t}\gamma$ and $W\gamma$+jets normalizations are consistent with the nominal prediction within the uncertainties of $^{+14\%}_{-13\%}$ and $^{+20\%}_{-17\%}$. The observed significance exceeds (is less than) $5\sigma$ when the 0fj SR ($\geq$1fj SR) is excluded from the fit. However, the inclusion of the 0fj SR significantly improves the precision of the measured signal cross sections.
 
\begin{figure}[h]
\centering
\subfloat[]{\includegraphics[width=0.33\textwidth]{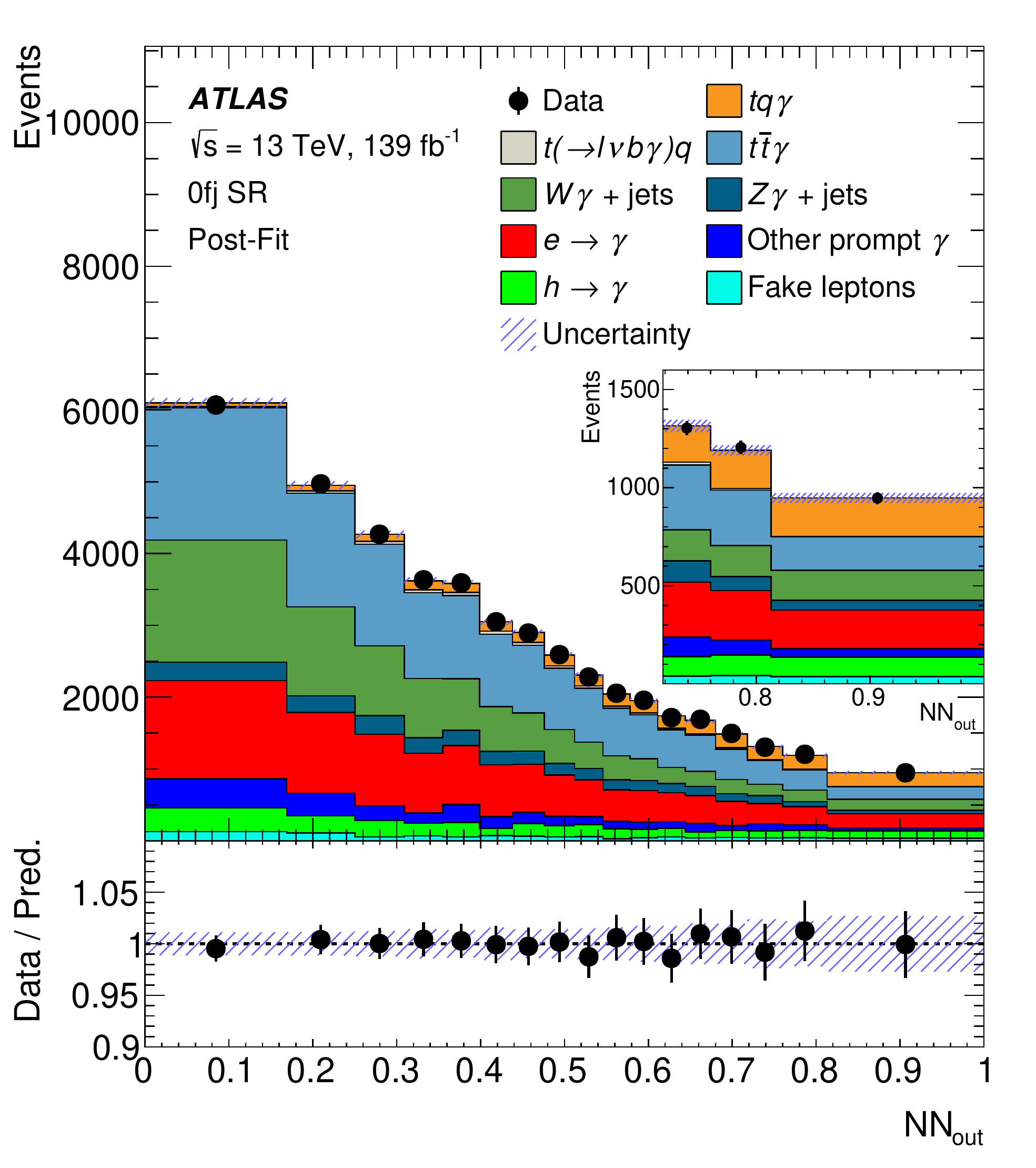}}
\subfloat[]{\includegraphics[width=0.33\textwidth]{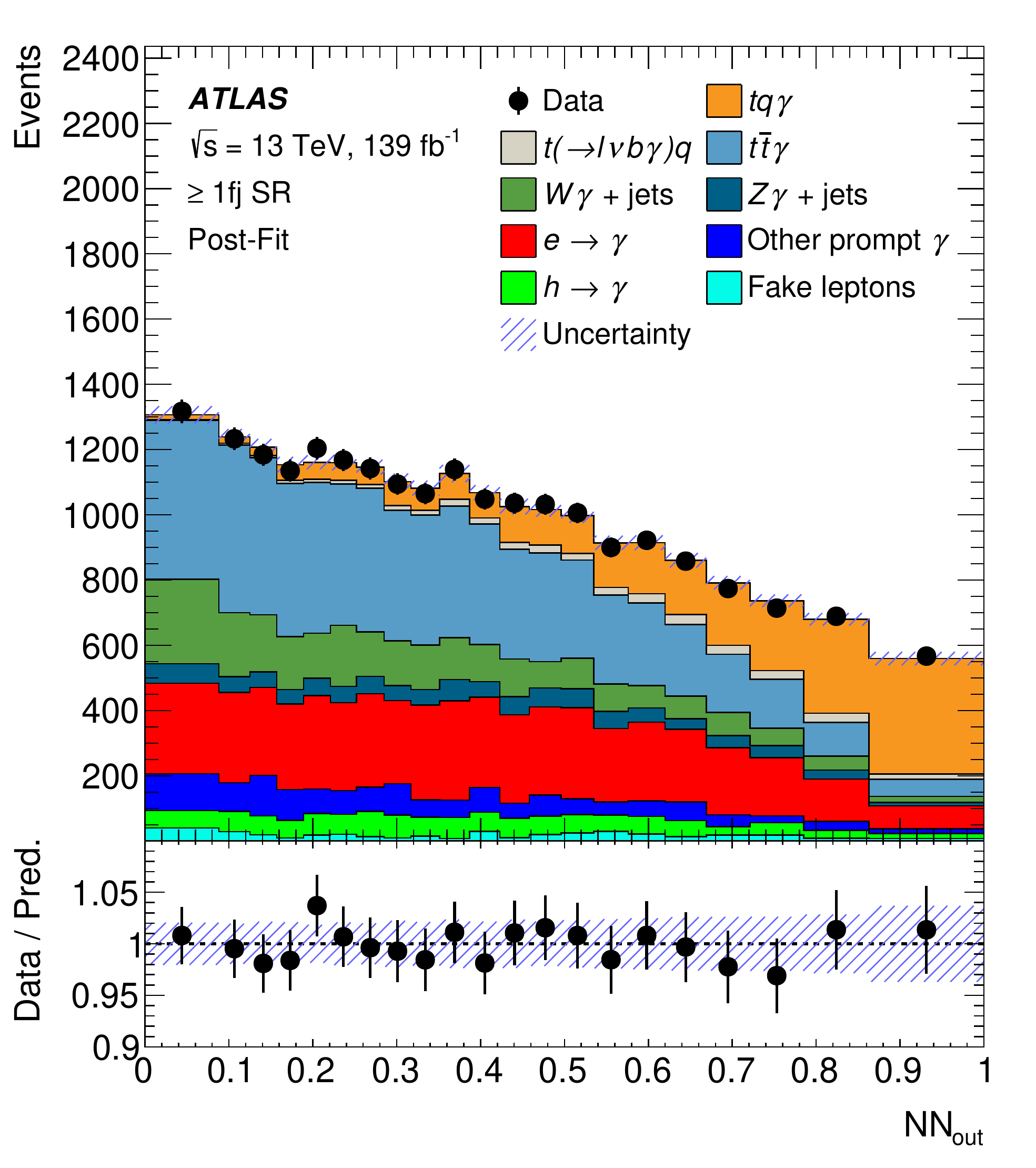}}
\subfloat[]{\includegraphics[width=0.33\textwidth]{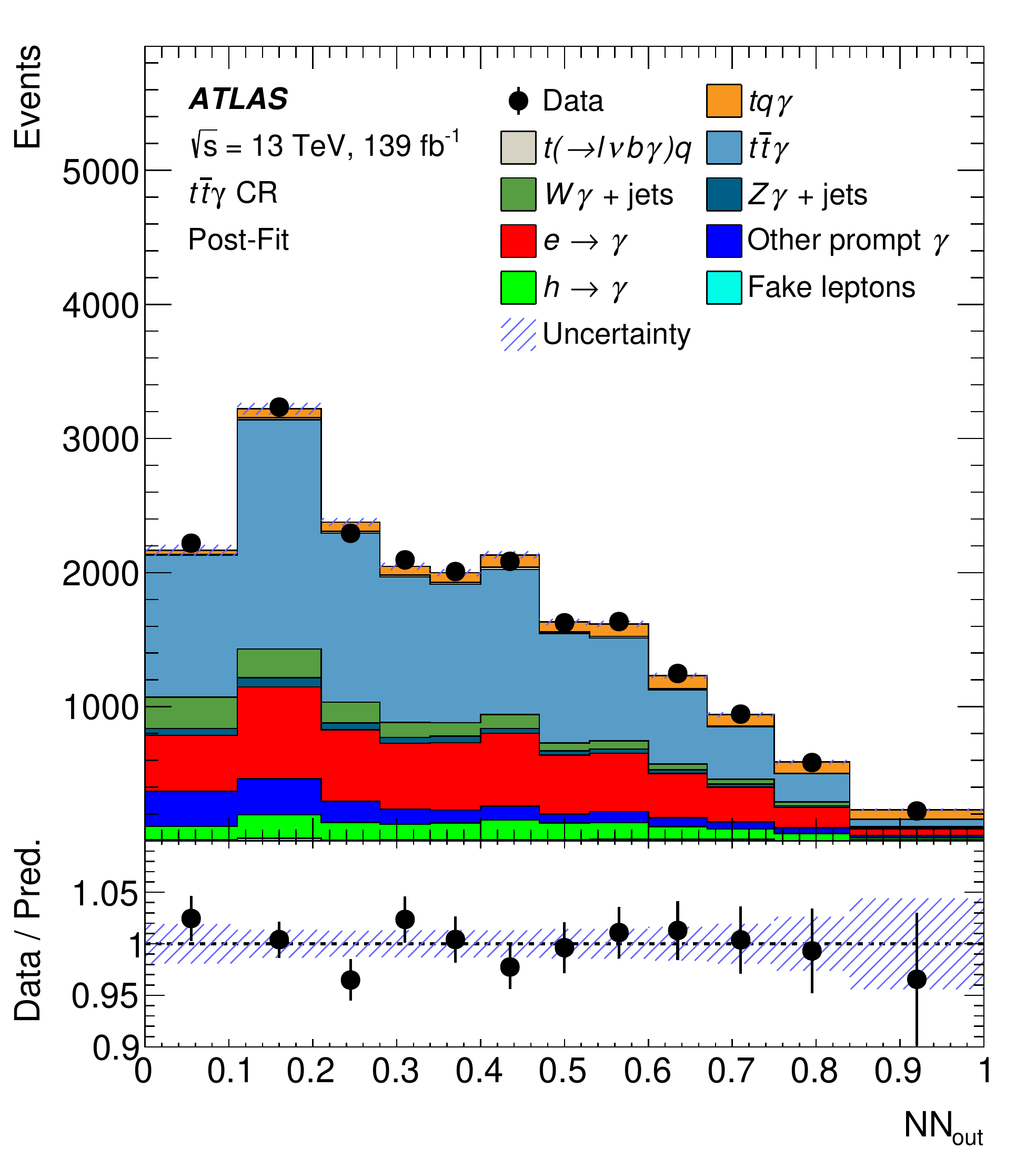}}
\caption{Distributions of the NN outputs in (a) the 0fj~SR, (b) the $\geq$1fj~SR, and (c) the $t\bar{t}\gamma$~CR in data and the expected contribution of the signal and background processes after the profile-likelihood fit. The hashed band represents the uncertainties in the SM prediction. The inset in (a) presents a close-up of the last three bins of the NN output.}
\label{fig:postfit}
\end{figure}
 
The measured fiducial cross section at stable-particle level is $\sigma_{\sigprod} \times \mathcal{B}\left(t\rightarrow \ell\nu b\right) + \sigma_{\sigdecay} = 303\,\pm\,9 \, \mathrm{(stat.)} \, ^{+33}_{-32}  \mathrm{(syst.)}$~fb. The measured fiducial cross section of $\sigprod$ production is $\sigma_{\sigprod}\times \mathcal{B}\left(t\rightarrow \ell\nu b\right) = 688\,\pm\,23 \, \mathrm{(stat.)} \, ^{+75}_{-71} \, \mathrm{(syst.)}$~fb. Both phase-space definitions require the photon $\pt$ to be at least $20~\GeV$. The precision of both measured fiducial cross sections of about 11\% is mainly limited by systematic uncertainties. The main sources of systematic uncertainty in the measurement of the fiducial cross section at stable-particle level (of $\sigprod$ production) are the modeling of $t\bar{t}\gamma$ production with $\pm$5.5\% ($\pm$5.5\%), the limited number of MC events for the background processes with $\pm$3.6\% ($\pm$3.5\%) and for the $\sigprod$ process with $\pm$3.0\% ($\pm$3.3\%), and the modeling of the $\sigdecay$ process with $\pm$3.3\% ($\pm$1.9\%) and of the $t\bar{t}$ process with $\pm$2.3\% ($\pm$2.4\%). The uncertainty in the modeling of the $\sigdecay$ process has a larger impact on the measurement at stable-particle level, because the $\sigdecay$ contribution is considered as part of the signal and is hence not fixed to the SM expectation.
 
The measured fiducial cross section at stable-particle level (of $\sigprod$ production) is compatible with the SM predictions at NLO in \alphas of $217\,^{+27}_{-15}$~fb ($515\,^{+36}_{-42}$~fb) within 2.0 (2.1) standard deviations. The 30\%--40\% higher measured cross sections are consistent with the results of the CMS measurement~\cite{CMS-TOP-17-016}, which yielded $1.42 \pm 0.43$ times the SM prediction in a slightly different fiducial phase space. It will need to be studied whether the small tension between measurements and theory predictions becomes more significant in future works on $\sigprod$ production. Corrections at approximate next-to-next-to-leading order (NNLO) in a similar phase space were found to enhance the $\sigprod$ cross section by 5.1\%~\cite{Kidonakis:2022ocq}. In conclusion, this Letter observed the associated production of a single top quark and a photon and measured its cross section in fiducial phase spaces with a precision of 11\%. It hence completes the picture of top-quark-associated production with the gauge bosons in the electroweak sector and establishes a new direct probe of the coupling of the top quark to the photon, which plays an important role to deepen the understanding of the top-quark's electroweak coupling.
 
\FloatBarrier


\section*{Acknowledgements}


We thank CERN for the very successful operation of the LHC, as well as the
support staff from our institutions without whom ATLAS could not be
operated efficiently.
 
We acknowledge the support of
ANPCyT, Argentina;
YerPhI, Armenia;
ARC, Australia;
BMWFW and FWF, Austria;
ANAS, Azerbaijan;
CNPq and FAPESP, Brazil;
NSERC, NRC and CFI, Canada;
CERN;
ANID, Chile;
CAS, MOST and NSFC, China;
Minciencias, Colombia;
MEYS CR, Czech Republic;
DNRF and DNSRC, Denmark;
IN2P3-CNRS and CEA-DRF/IRFU, France;
SRNSFG, Georgia;
BMBF, HGF and MPG, Germany;
GSRI, Greece;
RGC and Hong Kong SAR, China;
ISF and Benoziyo Center, Israel;
INFN, Italy;
MEXT and JSPS, Japan;
CNRST, Morocco;
NWO, Netherlands;
RCN, Norway;
MEiN, Poland;
FCT, Portugal;
MNE/IFA, Romania;
MESTD, Serbia;
MSSR, Slovakia;
ARRS and MIZ\v{S}, Slovenia;
DSI/NRF, South Africa;
MICINN, Spain;
SRC and Wallenberg Foundation, Sweden;
SERI, SNSF and Cantons of Bern and Geneva, Switzerland;
MOST, Taiwan;
TENMAK, T\"urkiye;
STFC, United Kingdom;
DOE and NSF, United States of America.
In addition, individual groups and members have received support from
BCKDF, CANARIE, Compute Canada and CRC, Canada;
PRIMUS 21/SCI/017 and UNCE SCI/013, Czech Republic;
COST, ERC, ERDF, Horizon 2020, ICSC-NextGenerationEU and Marie Sk{\l}odowska-Curie Actions, European Union;
Investissements d'Avenir Labex, Investissements d'Avenir Idex and ANR, France;
DFG and AvH Foundation, Germany;
Herakleitos, Thales and Aristeia programmes co-financed by EU-ESF and the Greek NSRF, Greece;
BSF-NSF and MINERVA, Israel;
Norwegian Financial Mechanism 2014-2021, Norway;
NCN and NAWA, Poland;
La Caixa Banking Foundation, CERCA Programme Generalitat de Catalunya and PROMETEO and GenT Programmes Generalitat Valenciana, Spain;
G\"{o}ran Gustafssons Stiftelse, Sweden;
The Royal Society and Leverhulme Trust, United Kingdom.
 
The crucial computing support from all WLCG partners is acknowledged gratefully, in particular from CERN, the ATLAS Tier-1 facilities at TRIUMF (Canada), NDGF (Denmark, Norway, Sweden), CC-IN2P3 (France), KIT/GridKA (Germany), INFN-CNAF (Italy), NL-T1 (Netherlands), PIC (Spain), ASGC (Taiwan), RAL (UK) and BNL (USA), the Tier-2 facilities worldwide and large non-WLCG resource providers. Major contributors of computing resources are listed in Ref.~\cite{ATL-SOFT-PUB-2023-001}.


\clearpage

\part*{Appendix: Systematic uncertainties}
 
This appendix contains additional information about signal and background MC samples and their systematic uncertainties.
 
In order to remove the overlap between the $\sigprod$ and $\sigdecay$ samples, events from the $\sigdecay$ sample are kept when the hypothesis of a radiative-decay photon better approximates the true $W$-boson or top-quark mass, i.e.\ either the $\ell\nu\gamma$ or the $\ell\nu b\gamma$ invariant mass is closer than the $\ell\nu$ or $\ell\nu b$ invariant mass to the $W$-boson or top-quark mass, respectively. The uncertainty associated with this procedure in the prediction of the fiducial cross section at stable-particle level is conservatively estimated to be 20\%, based on $\sigprod$ events that are falsely categorized as $\sigdecay$.
 
The background MC samples include the following production processes: photon radiation in $t\bar{t}$ production (NLO), $t\bar{t}$ with radiative decay (LO), $W\gamma/Z\gamma$+jets (NLO for up to one additional parton, LO for up to three)~\cite{Bothmann:2019yzt,Gleisberg:2008fv,Schumann:2007mg,Hoeche:2011fd,Hoeche:2012yf,Catani:2001cc,Hoeche:2009rj,Buccioni:2019sur,Cascioli:2011va,Buccioni:2017yxi,Denner:2016kdg}, $t\bar{t}$~\cite{Frixione:2007nw,Nason:2004rx,Frixione:2007vw,Alioli:2010xd}, single top quark~\cite{Frederix:2012dh,Frixione:2008yi}, $W$/$Z$+jets (NLO for up to two additional partons, LO for up to four), and diboson (NLO for up to one additional parton, LO for up to three). The overlap between samples with photons generated in the matrix element and those with photons from the parton shower is removed using generator-level information. The numbers of events from several MC samples are normalized to cross sections calculated to higher orders in \alphas: NNLO plus next-to-next-to-leading-logarithm precision for $t\bar{t}$ production~\cite{Beneke:2011mq,Cacciari:2011hy,Baernreuther:2012ws,Czakon:2012zr,Czakon:2012pz,Czakon:2013goa,Czakon:2011xx}, NNLO precision for $W$+jets and $Z$+jets production~\cite{Anastasiou:2003ds}, and NLO (NNLO) precision for single-top-quark production in the $t$ and $s$ channel~\cite{Kant:2014oha,Aliev:2010zk} ($tW$ channel~\cite{Kidonakis:2010ux}). For $t\bar{t}$ production with radiative decay, a LO-to-NLO correction factor of 1.67 is determined by subtracting the NLO \MGNLO prediction for the $t\bar{t}\gamma$ process from an NLO calculation of the full process~\cite{Melnikov:2011ta}.
 
The uncertainties in the inclusive cross sections amount to 6\% for $t\bar{t}$~\cite{Ball:2012cx,Czakon:2011xx,Butterworth:2015oua,Martin:2009bu,Gao:2013xoa}, 5.3\% for single-top-quark production~\cite{Kidonakis:2011wy,Kidonakis:2010tc,Kidonakis:2010ux}, 5\% for $W$+jets and $Z$+jets~\cite{ATL-PHYS-PUB-2016-003}, 30\% for $Z\gamma$+jets, and 50\% for diboson production, mostly in association with $b$ jets. An additional uncertainty of 30\% is assigned for the normalization of $W\gamma$ production in association with $b$ jets. The possible phase-space dependence of the LO-to-NLO correction factor for $t\bar{t}$ production with radiative decay is estimated by changing the correction factor from 1.67 to 1.97, motivated by the correction determined in Ref.~\cite{TOPQ-2017-14}. A 30\% uncertainty is assigned to the normalization of the $\sigdecay$ process, conservatively taken to be of the order of the difference between the predicted $\sigdecay$ event yields at LO and NLO. Uncertainties in the fake-lepton background arise from the uncertainties in the prompt-lepton subtraction in the matrix method and from a 50\% normalization uncertainty. The uncertainty in the integrated luminosity is 1.7\%~\cite{ATLAS-CONF-2019-021}. The uncertainty in the simulation of pileup is estimated by varying the average expected number of interactions per bunch crossing by 3\%.
 
Modeling uncertainties are evaluated as follows. Renormalization and factorization scales as well as PDFs are varied in the signal and background MC samples. The uncertainty from the choice of MC generator is estimated by comparing the nominal $\sigdecay$, $t\bar{t}$, and $tW$ samples with alternative samples generated with \MGNLO\ interfaced to \Pythia[8]. For the $tW$ sample, the difference between the diagram-subtraction scheme and the nominal diagram-removal scheme~\cite{Frixione:2008yi} is used as an uncertainty. The uncertainty from the choice of parton-shower program is estimated by comparing the nominal signal, $t\bar{t}\gamma$, $t\bar{t}$, and $tW$ samples with samples interfaced to \HERWIG[7]~\cite{Bahr:2008pv,Bellm:2015jjp}. The $t\bar{t}$ sample is compared with a sample with the value of the \hdamp parameter, controlling the $\pt$ of the first gluon emission in the \POWHEG\ generator, increased from $1.5\,m_{\mathrm{top}}$ to $3\,m_{\mathrm{top}}$~\cite{ATL-PHYS-PUB-2020-023}. The uncertainty in the modeling of initial- and final-state radiation is estimated by systematic variations in the A14 tune~\cite{ATL-PHYS-PUB-2014-021} in the signal, $t\bar{t}\gamma$ and $t\bar{t}$ samples. In addition, an uncertainty in the $\sigdecay$ sample is estimated by comparing the shapes predicted by the nominal sample with the shapes predicted by a LO sample with the decay $\sigdecay$ simulated directly in the hard process with \MGNLO\ using the NNPDF3.0 PDF set and interfaced with \Pythia[8].


\clearpage
 
\printbibliography
 
\clearpage
 
\begin{flushleft}
\hypersetup{urlcolor=black}
{\Large The ATLAS Collaboration}

\bigskip

\AtlasOrcid[0000-0002-6665-4934]{G.~Aad}$^\textrm{\scriptsize 101}$,
\AtlasOrcid[0000-0002-5888-2734]{B.~Abbott}$^\textrm{\scriptsize 119}$,
\AtlasOrcid[0000-0002-7248-3203]{D.C.~Abbott}$^\textrm{\scriptsize 102}$,
\AtlasOrcid[0000-0002-1002-1652]{K.~Abeling}$^\textrm{\scriptsize 55}$,
\AtlasOrcid[0000-0002-8496-9294]{S.H.~Abidi}$^\textrm{\scriptsize 29}$,
\AtlasOrcid[0000-0002-9987-2292]{A.~Aboulhorma}$^\textrm{\scriptsize 35e}$,
\AtlasOrcid[0000-0001-5329-6640]{H.~Abramowicz}$^\textrm{\scriptsize 150}$,
\AtlasOrcid[0000-0002-1599-2896]{H.~Abreu}$^\textrm{\scriptsize 149}$,
\AtlasOrcid[0000-0003-0403-3697]{Y.~Abulaiti}$^\textrm{\scriptsize 116}$,
\AtlasOrcid[0000-0003-0762-7204]{A.C.~Abusleme~Hoffman}$^\textrm{\scriptsize 136a}$
% ...
\bigskip
\\

% ...
$^{29}$Physics Department, Brookhaven National Laboratory, Upton NY; United States of America.\\
% ...
$^{35}$$^{(a)}$Facult\'e des Sciences Ain Chock, R\'eseau Universitaire de Physique des Hautes Energies - Universit\'e Hassan II, Casablanca;$^{(b)}$Facult\'{e} des Sciences, Universit\'{e} Ibn-Tofail, K\'{e}nitra;$^{(c)}$Facult\'e des Sciences Semlalia, Universit\'e Cadi Ayyad, LPHEA-Marrakech;$^{(d)}$LPMR, Facult\'e des Sciences, Universit\'e Mohamed Premier, Oujda;$^{(e)}$Facult\'e des sciences, Universit\'e Mohammed V, Rabat;$^{(f)}$Institute of Applied Physics, Mohammed VI Polytechnic University, Ben Guerir; Morocco.\\
% ...
$^{55}$II. Physikalisches Institut, Georg-August-Universit\"{a}t G\"ottingen, G\"ottingen; Germany.\\
% ...
$^{101}$CPPM, Aix-Marseille Universit\'e, CNRS/IN2P3, Marseille; France.\\
$^{102}$Department of Physics, University of Massachusetts, Amherst MA; United States of America.\\
% ...
$^{116}$Department of Physics, New York University, New York NY; United States of America.\\
% ...
$^{119}$Homer L. Dodge Department of Physics and Astronomy, University of Oklahoma, Norman OK; United States of America.\\
% ...
$^{136}$$^{(a)}$Departamento de F\'isica, Pontificia Universidad Cat\'olica de Chile, Santiago;$^{(b)}$Millennium Institute for Subatomic physics at high energy frontier (SAPHIR), Santiago;$^{(c)}$Instituto de Investigaci\'on Multidisciplinario en Ciencia y Tecnolog\'ia, y Departamento de F\'isica, Universidad de La Serena;$^{(d)}$Universidad Andres Bello, Department of Physics, Santiago;$^{(e)}$Instituto de Alta Investigaci\'on, Universidad de Tarapac\'a, Arica;$^{(f)}$Departamento de F\'isica, Universidad T\'ecnica Federico Santa Mar\'ia, Valpara\'iso; Chile.\\
% ...
$^{149}$Department of Physics, Technion, Israel Institute of Technology, Haifa; Israel.\\
$^{150}$Raymond and Beverly Sackler School of Physics and Astronomy, Tel Aviv University, Tel Aviv; Israel.\\
% ...
$^{a}$ Also Affiliated with an institute covered by a cooperation agreement with CERN.\\
$^{b}$ Also at An-Najah National University, Nablus; Palestine.\\
$^{c}$ Also at Borough of Manhattan Community College, City University of New York, New York NY; United States of America.\\
$^{d}$ Also at Bruno Kessler Foundation, Trento; Italy.\\
$^{e}$ Also at Center for High Energy Physics, Peking University; China.\\
$^{f}$ Also at Centro Studi e Ricerche Enrico Fermi; Italy.\\
% ...
$^{t}$ Also at Department of Physics, Westmont College, Santa Barbara; United States of America.\\
% ...

\end{flushleft}


\end{document}